\documentclass[12pt]{article}  
 
\usepackage{epsfig} 
\usepackage{amsmath,amssymb,amsfonts} 
\usepackage{float,a4wide} 
\newcommand{\be}{\begin{equation}} 
\newcommand{\en}{\end{equation}}
\newcommand{\bea}{\begin{eqnarray}}
\newcommand{\ena}{\end{eqnarray}}
\newcommand{\Det}{\hbox{Det}}
\newcommand{\hbo}{\hbox to 1 true cm {\hfill } }

\def\dslash{\partial\kern-.5em\slash}
\def\kslash{k\kern-.5em\slash}
\def\pslash{p\kern-.5em\slash}
\def\Dslash{D\kern-.5em\slash}

\begin{document} 
\vglue 1truecm
  
\vbox{ 
\hfill January 9, 2003
}
  
\vfil
\centerline{\large\bf The Gribov problem and its solution from } 
\centerline{\large\bf a toy model point of view }

\bigskip
\centerline{ Kurt Langfeld } 
\bigskip
\vspace{.5 true cm} 
\centerline{ Institut f\"ur Theoretische Physik, Universit\"at 
   T\"ubingen }
\centerline{D--72076 T\"ubingen, Germany.}
\bigskip
\bigskip\bigskip

\vskip 1.5cm

\begin{abstract}
The standard Faddeev Popov gauge fixing procedure is put on 
solid grounds by taking into account a topological factor which
corrects for the number of Gribov copies within the first 
Gribov horizon. Zwanziger's stochastic approach to gauge fixed 
Yang-Mills theory is briefly reviewed. 
A simple toy model is presented which illustrates 
both methods. Within the toy model, I show that a stochastic drift 
force can be constructed with which the gauged configurations are 
attracted by the fundamental modular region. The toy model 
shows that an action which gives rise to the drift force can be
found. This makes a ``heat bath'' simulation possible, 
which is seen to be superior to the Langevin approach at the numerical 
level.

\end{abstract}

\vfil
\hrule width 5truecm
\vskip .2truecm
\begin{quote} 

keywords: Gribov copies, stochastic quantization, fundamental modular 
region, Gribov horizon
\end{quote}
\eject

\section{ Introduction }
\label{sec:1} 

Although lattice gauge simulations can directly address gauge
invariant, physical observables of the theory of strong interactions, 
it might turn out fruitful for detecting basic mechanisms 
to remove redundant degrees of freedom by means of gauge fixing. 
Moreover, in the case of the perturbative approach to QCD Greenfunctions 
as well as in the case employing 
Dyson-Schwinger techniques~\cite{Alkofer:2000wg,Roberts:2000aa}, 
gauge fixing is inevitable.

\vskip 0.3cm 
In the Faddeev Popov approach, 
gauge fixing is imposed by demanding that the gauge 
fields obey the gauge fixing condition. The 
degeneracy factor which determines the weight with which a 
representative of the gauge orbit contributes to the 
partition function is assumed to be given by the Faddeev Popov 
determinant. It was firstly pointed out that e.g.~the well known
Landau gauge condition is not sufficient for an unambiguous choice of 
the gauge field~\cite{Gribov:1977wm}. As a further refinement of the 
gauge condition, one might demand that the lowest eigenvalue of the 
Faddeev Popov matrix evaluated with gauged fields, is 
positive~\cite{Gribov:1977wm,Zwanziger:1981kg}.
These gauge fixed 
configurations are said to lie within the first Gribov horizon. 
It still turns out that more than one representative of the gauge 
orbit lies within the first Gribov regime in the generic
(non-perturbative) case~\cite{vanBaal:1991zw}. 
Finally, the domain of gauged 
configurations which have been unambiguously selected from the first 
Gribov regime is known as the fundamental modular region. 
In practice, any of the refinements of the naive Landau gauge
condition can be used for a calculation of the gauge invariant 
observables, once the weight factor of each configuration is known. 
The Gribov problem emerges when this weight factor is calculated 
by the standard Faddeev Popov method: it was pointed out 
in~\cite{Baulieu:1996kb,Baulieu:1996rp,Schaden:1998hz} that 
the inverse Faddeev Popov determinant identically vanishes leaving us 
with an ill-defined partition function. There are two possibilities 
which are actually used in lattice gauge theories to avoid 
this problem: (i) gauge fixed configurations are randomly selected 
from the first Gribov regime, and their weight factor is 
exactly taken into account (see 
e.g.~\cite{Mandula:1998tt}-\cite{Langfeld:2001cz}); this method does 
not suffer from the Faddeev-Popov problem, but the selection 
of the configurations from the first Gribov regime is still
arbitrary. This could lead to ambiguous results for gauge variant 
quantities, such as the gluon propagator; (ii) a Laplacian 
version~\cite{Vink:1992ys} of the gauge condition is chosen which 
allows a unique definition of the gauged configuration 
in practical simulations (see e.g.~\cite{Alexandrou:2000ja}). 

\vskip 0.3cm 
In order to avoid the Gribov problem in an ab initio continuum 
formulation of Yang-Mills theory, it was proposed by  
Zwanziger~\cite{Zwanziger:1981kg,Zwanziger:2002ia} to select 
stochastically, but in 
a well defined manner the gauge field configurations from the 
configuration space. Thereby, preference is given to configurations 
within or close to the first Gribov regime. Using the framework of 
the stochastic quantization~\cite{parisi}, the bias towards the 
first Gribov regime is provided by a drift force ``tangent'' to 
the gauge orbit. 

\vskip 0.3cm 
In the present paper, I briefly review the gauge fixing techniques 
of continuum Yang-Mills theory. A simple toy model is used 
to illustrate the Gribov problem of the standard Faddeev Popov 
quantization, and its resolution in the stochastic approach. 
I find in the case of the toy model that a modification of the 
stochastic approach generates configurations which are attracted 
by the fundamental modular region. I finally point out that a 
``potential'' representation of the drift force seems possible. 
This paves the path to a very efficient numerical simulation 
using the heat bath techniques.

\section{ Brief review of gauge fixing }

\subsection{ The standard gauge fixing procedure }
\label{sec:1.1} 

The task boils down to select a single gauge field 
configuration of $\{ A_\mu ^\Omega (x)\}$, where $\{ A_\mu ^\Omega
(x)\}$ is the set of gauge fields $A_\mu ^\Omega (x)$ which have 
been generated from a representative $A_\mu(x)$ of the gauge orbit 
by applying all possible 
gauge transformations $\Omega (x)$.  Usually, a gauge fixing 
condition, e.g. such as Landau gauge 
\be 
\Omega (x) \; : \hbo \partial _\mu A_\mu ^\Omega (x) \; = \; 0 \; , 
\label{eq:i1}
\en
is chosen to accomplish this task. Unfortunately, this procedure is 
plagued by the so-called Gribov problem~\cite{Gribov:1977wm}: 
background fields $A_\mu (x)$ which give rise to ``large'' field strength 
generically admit several solutions to the gauge fixing 
condition (for an illustration see~\cite{Bruckmann:2000xd}). 
This implies that the topological field theory, which generates to 
the inverse Faddeev-Popov determinant, 
vanishes identically~\cite{Baulieu:1996kb,Baulieu:1996rp,Schaden:1998hz}
\be 
\Delta ^{-1}_{FP} \; = \; \int {\cal D} \Omega \; \delta \Bigl( 
\partial _\mu A_\mu ^\Omega (x) \Bigr) \; \equiv \; 0 \; .
\label{eq:FP} 
\en
In the present paper, a lattice regularization as well as a finite 
space-time volume is implicitly understood in order to render 
functional integrals well defined. 
In order to select a gauge configuration $A^\Omega _\mu (x)$ from 
the possible solutions of (\ref{eq:i1}), one might demand 
\be 
\hbo  f[A^\Omega] := \partial _\mu A_\mu ^\Omega (x) \; = \;
0 \; , \hbo  {\cal M} := \frac{ \delta f[A^\Omega] }{ \delta \Omega }
\; , \hbo 
\lambda _0 > 0 \; , 
\label{eq:i2}
\en
where $\lambda _0$ is lowest eigenvalue of the operator ${\cal M}$. 
The configurations $A^\Omega _\mu (x)$ emerging from the latter 
equation are said to be restricted to the first Gribov regime. 
It turns out, that even the refinement (\ref{eq:i2}) of gauge fixing does  
not yield a unique representative $A^\Omega _\mu (x)$ of the gauge
orbit. In order to count the number $n[A]$ of configurations 
lying within the first Gribov horizon, one evaluates the functional 
integral, i.e., 
$$ 
n[A]  \; = \; \int {\cal D} \Omega \; \; \delta \Bigl( 
\partial _\mu A_\mu ^\Omega (x) \Bigr) \; \; 
\Det \Bigl( \frac{ \delta f[A^\Omega] }{ \delta \Omega }
\Bigr) \; \;  \theta \Bigl[ \lambda _0 \Bigr] \; , 
$$
where the modulus of the functional determinant has been replaced by 
the determinant in view of the constraint to positive eigenvalues only. 
Since the gauge condition (\ref{eq:i2}) can be installed using the 
variational condition 
$$ 
\int d^4x \; A_\mu^2  (x) \; \stackrel{\Omega }{ \rightarrow } \; 
\hbox{minimum} \; , 
$$ 
it is clear that at least one representative can be found on the 
gauge orbit implying that $n[A] \ge 1$. Hence, the partition function 
of gauged fixed Yang-Mills theory is well defined and is given by 
\be
Z \; = \;  \int {\cal D} A \; n^{-1}[A] \; \delta \Bigl( 
\partial _\mu A_\mu  \Bigr) \; \; 
\Det \Bigl( \frac{ \delta f[A] }{ \delta \Omega }
\Bigr) \; \; 
\theta \Bigl[ \lambda _0 \Bigr] \; \exp \Bigl\{ - S_{\mathrm YM} \Bigr\} \; . 
\label{eq:a3} 
\en
The determinant in the latter equation can be represented by 
ghost fields as usual. 
At the current stage of investigations present in the literature 
the topological factor $n[A]$ is set to unity. A perturbative
treatment of the partition
function  is consistent with the approximation $n[A]=1$, since only 
``small'' field strengths are involved by construction. 
The influence of the  factor $n[A]$ on a genuine non-perturbative
approach is not known. To my knowledge, an analytic expression of the 
integer $n[A]$ is not available. 

\vskip 0.3cm
In order to circumvent these difficulties, Baulieu and Schaden 
proposed~\cite{Baulieu:1996kb}-\cite{Schaden:1998hz} to perform 
subsequent steps of gauge fixing where each of them avoids 
the topological obstruction (\ref{eq:FP}). 

\subsection{ The stochastic approach to gauge fixed YM-theory }
\label{sec:1.2} 

It was pointed by Zwanziger~\cite{Zwanziger:1981kg,Zwanziger:2002ia} that 
the Gribov problem can be avoided when stochastic quantization~\cite{parisi}
is used in a modified form: a Langevin simulation generates a series of 
configurations $\{A_\mu (x)\}^L$ where each configuration possesses a 
certain bias concerning its position on the gauge orbit. At the same
time, the correct result for gauge invariant observables is 
recovered when the observable is calculated from the Langevin ``time''
series. In order to be specific, let me firstly point out
that Langevin time series is generated by 
\be 
\frac{\partial }{ \partial \tau } A_\mu (x) \; = \; - \; \frac{ 
\delta S_{YM} }{ \delta A_\mu (x) } \; + \; {\cal K} _\mu (x) 
\; + \; [\hbox{noise}] \; , 
\label{eq:i10} 
\en 
where $S_{YM}$ is the Yang-Mills action, and ${\cal K} _\mu (x) $ 
is a drift force, which will be specified later. Expectations values 
and Greenfunctions are evaluated using~\cite{parisi} 
\be 
\int {\cal D} A_\mu \; {\cal G}[A] \; P[A] 
\; \propto \; \lim _{T \to \infty } \frac{1}{T} \int _0^T \; 
d\tau \; {\cal G} \Bigl[ A_\mu (x; \tau ) \Bigr] \; . 
\label{eq:i11} 
\en
The probabilistic weight $P[A]$ satisfies a Fokker-Planck 
equation
\be 
H_{FP} \, P[A] \; = \; 
\int d^4 x \; \frac{ \delta }{ \delta A_\mu (x) } \; \Bigl[ 
- \frac{ \delta P[A] }{ \delta A_\mu (x) } \; + \; \Bigl( 
\, - \, \frac{ \delta S_{YM} }{ \delta A_\mu (x) } \; + \; 
 {\cal K} _\mu (x) \Bigl) \; P[A] \; \Bigl] \; = \; 0 \; .
\label{eq:i12} 
\en
For a vanishing Drift force, i.e., ${\cal K} _\mu (x)=0$, one recovers 
the desired result 
$$
P[A] \; \propto \; \exp \Bigl\{ - S_{YM} \Bigr\} \; . 
$$

The key observation~\cite{Zwanziger:2002ia}  made by Zwanziger
is that if one chooses 
a drift force ``tangent'' to the gauge orbit, the gauge fields 
are generated with preference to a certain region of the 
gauge orbit, and, hence, are interpreted as gauge fields of a certain 
gauge. In the context of Landau gauge, one chooses 
$$
{\cal K} _\mu \; = \; D_\mu \, v(x) \; , 
$$ 
where $D_\mu $ is the gauge covariant derivative and $v(x)$ is an 
arbitrary auxiliary field~\cite{Zwanziger:2002ia}. 
On the other hand, gauge invariant observables are independent of 
${\cal K} _\mu (x)$, and, therefore, the right hand side of
(\ref{eq:i11}) reproduces the correct Yang-Mills result for 
gauge invariant quantities. The goal of the work~\cite{Zwanziger:2002ia} 
is that a theoretical, Dyson-Schwinger type framework is developed 
for the stochastic approach. A numerical analysis using the 
stochastic approach can be found in~\cite{Mizutani:za,Aiso:au}.

\section{  The toy model }
\label{sec:2} 

\subsection{ Settings }
\label{sec:2.1} 

Instead of a functional integral over the gauge fields $A_\mu (x)$, 
I will study a simple integral of two variables $x_1=x$, $x_2=y$. 
Gauge invariance is replaced by the rotational invariance in 
two dimensions. The ``gauge invariant'' action is given 
by 
\be 
S_{YM} \; = \; x^2 + y^2 \; = x_k^2 \; . 
\label{eq:1} 
\en 
In this very simple toy model, a separation of the degrees of freedom 
into gauge invariant and gauge dependent ones is done by 
introducing polar coordinates $(r, \varphi )$. ``Gauge invariant 
observables'' are functions of $r$ only. We are interested in 
expectation values of ``gauge invariant'' observables: 
\be 
\Bigl\langle f(r) \Bigr\rangle \; = \; \frac{1}{N} 
\int dx \; dy \; f(r) \; \exp \Bigl\{ - S_{YM} \Bigr\}\; ,  
\label{eq:2} 
\en 
where $r:=\sqrt{x^2+y^2}$. 
If $\vec{x}_0$ denotes a representative of a gauge orbit, 
the orbit $\{\vec{x}^\Omega \}$ is generated by 
\be 
\vec{x}^ \Omega \; = \; \left( \begin{array}{cc}
\cos \theta & - \sin \theta \cr 
\sin \theta &  \cos \theta \cr  \end{array} \right) \; \vec{x}_0 \; , 
\hbo \theta \in [0, 2\pi [ \; . 
\label{eq:i20} 
\en

\subsection{ Standard gauge fixing } 

In order to remove the rotational degree of freedom, we introduce 
a gauge fixing condition 
\be 
y(\theta ) \; = \; 0 \; .
\label{eq:i21} 
\en 
One easily verifies that the procedure suffers from the topological 
obstruction (\ref{eq:FP}), i.e., 
\be 
\Delta ^{-1}_{FP} \; = \; \int _0 ^{2 \pi } d \theta \; 
\delta \Bigl( y(\theta) \Bigr) \; = \; \int _0 ^{2 \pi } d \theta \; 
\delta \Bigl( \sin \theta \, x_0 \; + \; \cos \theta \; y_0  \Bigr) \; =
\; 0 
\label{eq:i22} 
\en 
independent of the gauge orbit specified by $(x_0,y_0)$. To simplify
the discussion below, we choose $x_0 \ge 0$, $y_0=0$ without a loss of 
generality. In order 
to refine the gauge fixing procedure according to (\ref{eq:a3}), 
we calculate 
\be 
 \Biggl[ \Det \Bigl( \frac{ \delta f[A] }{ \delta \Omega }
\Bigr) \Biggr] \; \rightarrow \; 
\frac{ d y(\theta ) }{ d \theta } \; = \; \cos \theta \, x_0 \; . 
\label{eq:i23} 
\en 
From 
$$ 
\frac{ d y(\theta ) }{ d \theta } \; = \; \cos \theta \; x_0 \; \ge \;
0 \; , 
$$ 
we conclude that the first Gribov regime consists of the positive 
half-space $x\ge 0$. Consequently, the topological quantity $n[A]$ 
becomes 
\be 
n[A] \; \rightarrow \; \; \; 
n [\vec{x}] \; = \; \int _{- \pi/2 }^{\pi/2} \, d\theta \; 
\cos \theta \; \delta \Bigr( \sin \theta \; x_0 \Bigr) \; = \; 1 \; . 
\label{eq:i24} 
\en 
In the case of the example (\ref{eq:i21}), there is only one
configuration within the first Gribov horizon. In this case, the 
first Gribov regime coincides with the 
fundamental modular region. 

\begin{figure}[t]
\centerline{  
\epsfxsize=7cm 
\epsffile{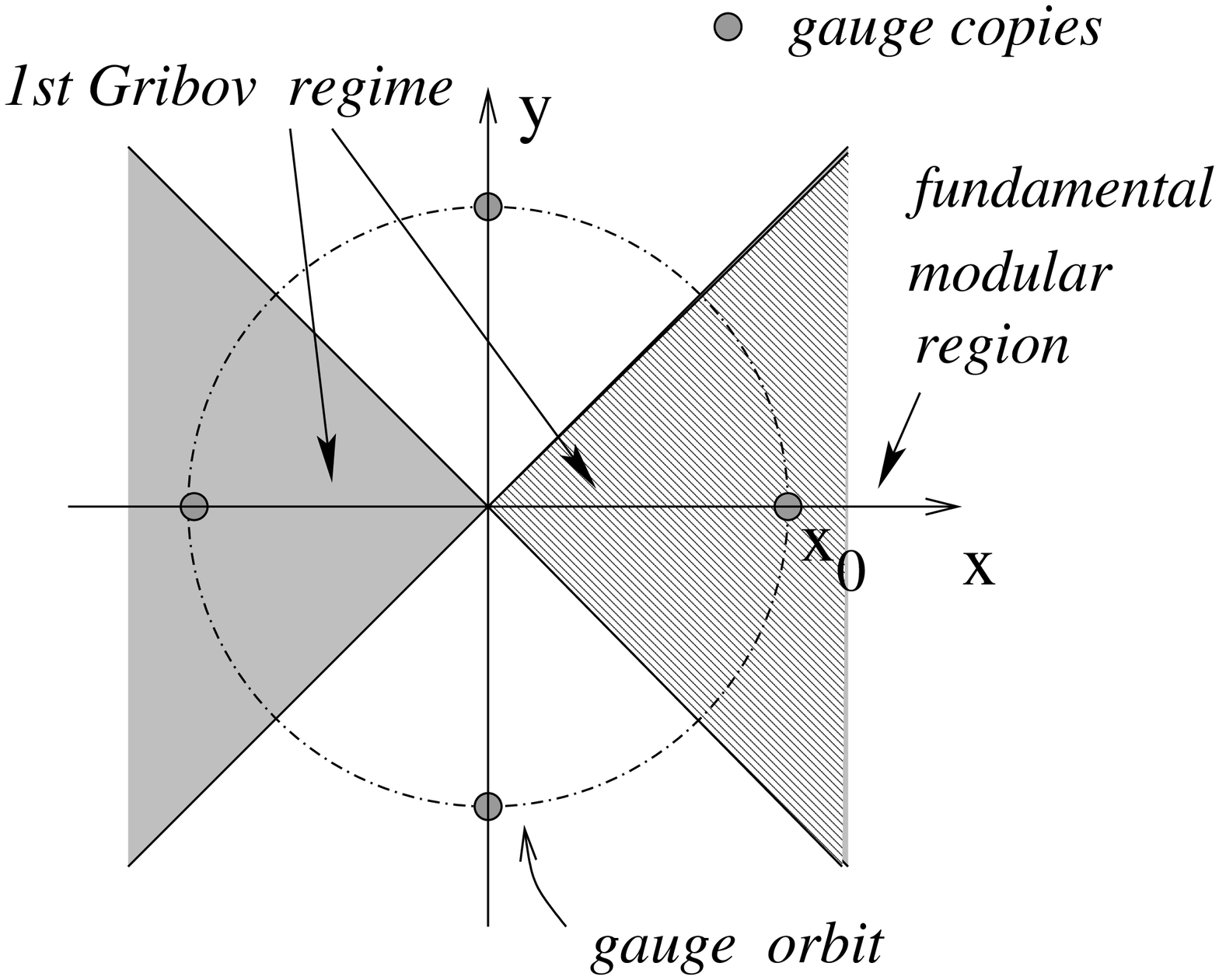} 
\epsfxsize=6.5cm 
\epsffile{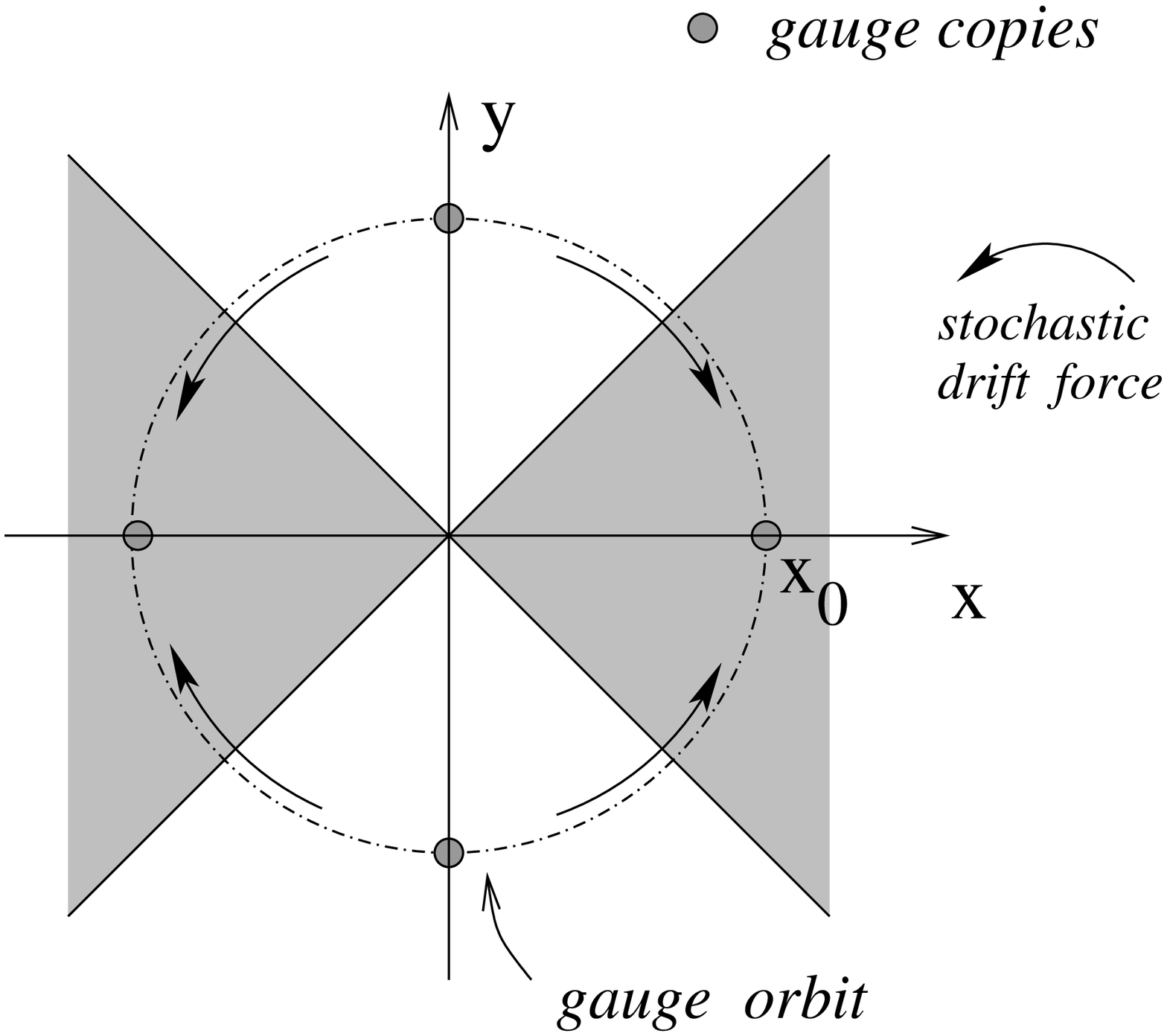} 
} 
\vskip 0.3cm
\caption{ Illustrations of the configuration space using the 
``gauge condition'' (\ref{eq:i25}). 
}
\label{fig:0}
\end{figure}

\vskip 0.3cm 
Let us study the more advanced example of a gauge fixing condition: 
\be 
f(\theta ) \; = \; x(\theta ) \; y(\theta ) \; = \; 0 \; , \rightarrow \; 
x_0^2 \; \sin \theta \; \cos \theta \; = \; \frac{1}{2} x_0^2 \; 
\sin (2 \theta ) \; = \; 0 \;. 
\label{eq:i25} 
\en 
In this case, four gauge copies exist. The first Gribov regime is 
determined from 
\be 
\frac{df(\theta )}{d\theta } \; = \; 
x_0^2 \; \cos (2 \theta ) \; \ge \; 0 \; , \hbo \theta \in [0, 2 \pi [ 
\; . 
\label{eq:i26} 
\en 
The space where (\ref{eq:i26}) holds is shaded in figure \ref{fig:0}. 
Hence, the first Gribov regime consists of the complete $x$-axis. 
A direct calculation of $n[\vec{x}]$ shows that there are still two 
gauge copies within the first Gribov horizon (see figure 
\ref{fig:0}). Nevertheless, we arrived 
at a stage where calculations of ``gauge invariant'' quantities 
are feasible: using $n[\vec{x}]=2$ and adapting (\ref{eq:a3}), we find 
\be 
Z \; = \; 2\pi \, 
\int dx \; dy \; \frac{1}{2} \; \delta \Bigl( x \, y \Bigr) 
\; \theta \Bigl[ x^2 - y^2 \Bigr] \; \exp \Bigl\{ - x^2 - y^2 
\Bigr\} \; . 
\; . 
\label{eq:i27} 
\en 
The result that $n[\vec{x}]$ is independent of $x_0$ is due to 
an oversimplification by the toy model. One can easily construct 
examples for the generic case, i.e., $n[\vec{x}]$ is a function of 
$x_0$.

\subsection{ The stochastic approach } 

In the present case of only two degrees of freedom, the 
gauge invariant part of the Fokker-Planck 
Hamiltonian (\ref{eq:i12}) is given by a partial differential 
equation 
\be 
H_{inv} \; = \; \frac{ \partial }{ \partial x_i } 
\biggl[ - \frac{ \partial  }{ \partial x_i }  \; + \; 
\Bigl( - \frac{ \partial S_{YM} }{ \partial x_i }  \Bigr) 
\; \biggr] \; . 
\label{eq:4} 
\en 
Infinitesimal rotations are generated by the operator 
\be 
G \; = \; x_i \; \epsilon _{ik} \, \frac{ \partial }{ \partial x_k }
\; ,
\label{eq:3} 
\en 
where $\epsilon _{ik}$ is the total anti-symmetric tensor in
2-dimensions. It is easy to check that the operator $G$ (\ref{eq:3}) 
commutes with $ H_{inv}$. Following the procedure suggested by
Zwanziger~\cite{Zwanziger:2002ia}, the drift force ``tangent to the 
gauge orbit'' is chosen as 
$K_i \; = \; \epsilon _{ik} x_k \, v(x) $, where the scalar function 
$v(x)$ will be specified below. 
The Fokker-Planck equation (which determines the weight factor $P_v$)
with the restoring drift force included is given by 
\be 
H_{FP} P_v \; = \; \frac{ \partial }{ \partial x_i } 
\biggl[ - \frac{ \partial P_v }{ \partial x_i }  \; + \; 
\Bigl( - \frac{ \partial S_{YM} }{ \partial x_i } \, + \,  
\epsilon _{ik} \, x_k \, v(x) \Bigr) P _v 
\; \biggr] \; = \; 0 \; . 
\label{eq:5} 
\en 

\vskip 0.3cm
{\bf Case study I:  The trivial case. } 

Let us study the ``gauge invariant'' scalar function $v(x) =
a^{-1}/r^2$. 
Using this particular choice, the Fokker-Planck equation
is given by
\be 
\biggl[ - \partial ^2 \; - \; 2 \, \partial _k x_k  \; + \; 
\frac{ a^{-1} }{r^2} \; \Bigl(
\, x_2 \, \partial _1 \, - x_1 \, \partial _2 \, \Bigr) \, \biggr] 
\; P_v \; = \; 0 \; . 
\label{eq:6a} 
\en 
To shed light onto the solution of (\ref{eq:6a}), we introduce 
polar coordinates: 
\be 
x_1=r \, \cos \varphi \, , \hbo 
x_2=r \, \sin \varphi \, , \hbo 
\varphi \in [0,2 \pi [ \; . 
\label{eq:coor} 
\en
One finds: 
\be 
\biggl[ - \partial _r ^2 \, - \,\frac{1}{r} \partial _r 
\; - \; \frac{1}{r^2} \, L^2 \; - \; 2 r \, \partial _r \; - \; 4 
\; + \; \frac{1}{a \, r^2 } \, L_3 \; \biggr] 
\; P_v \; = \; 0 \; ,  
\label{eq:6b} 
\en 
where 
\be 
L^2 \; = \; \partial _\varphi ^2 \; \hbo 
L_3 \; = \; \partial _\varphi \; . 
\label{eq:6c} 
\en 
As in quantum mechanics, the wave function $P_v$ factorizes (for this 
particular choice of $v(r)$): 
\be 
P_v \; = \; e^{ i m \, \varphi } \, \bar{P}_v(r) \; , \hbo 
m^2 \; + \; i \, a^{-1} m \; = \; 0 \; . 
\en
The only solution obeying periodic boundary conditions, i.e., 
$P_v(\varphi + 2 \pi) = P_v(\varphi )$, is $m=0$. 
The Fokker-Planck solution is hence given by 
\be 
P_v \; = \; 
\exp \Bigl\{ -r^2 \Bigr\} \; . 
\label{eq:6cc} 
\en 
One recovers the trivial result of the ``un-fixed'' action. 

\begin{figure}[t]
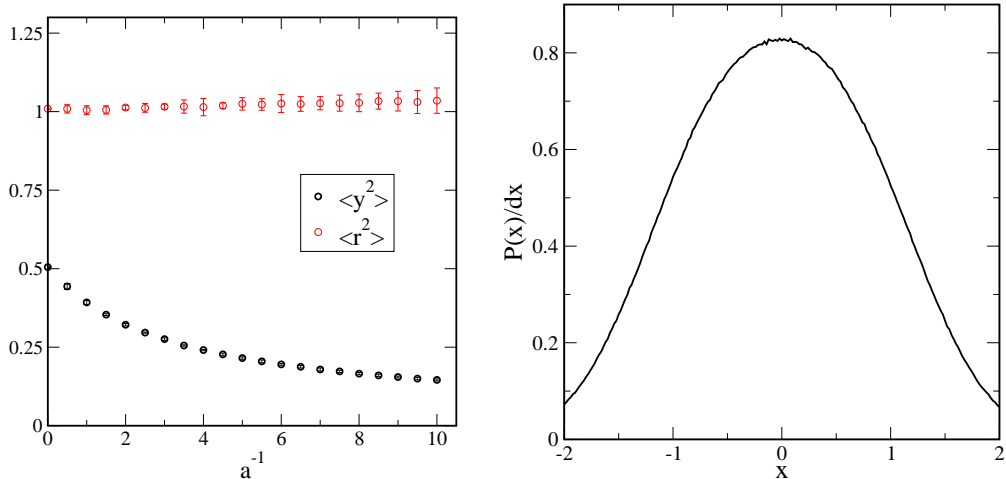

\centerline{  
\epsfxsize=6cm 
\epsffile{all.eps} 
\hspace{0.3cm}
\epsfxsize=6.7cm 
\epsffile{lanx.eps} 
} 
\vskip 0.3cm
\caption{ The expectation values $\langle r^2 \rangle $ 
   and $\langle y^2 \rangle $ 
   as function of the Landau parameter $a^{-1}$ (left panel). 
   The probability distribution of $x_n$ for $a^{-1}=5$ (right panel).
   Landau type drift force, see (\ref{eq:6d}).   
}
\label{fig:1}
\end{figure}

\vskip 0.3cm
{\bf Case study II. The Landau gauge. } 

Let us now investigate a drift force which is not constant 
along the gauge orbit, i.e., 
the ``Landau type gauge'' is given by the condition $x_1^\theta
x_2^\theta  = 0$
where the vector $x_k^\theta $ is generated from the vector 
$x_k $ by a rotation by an angle $\theta $ (see (\ref{eq:i20})). 
Here we choose the 
scalar function of the drift force according $v(x,y) = a^{-1} \, 
xy$, i.e., 
\be 
K_i \; = \; \epsilon _{ik} x_k \, v(x) \; , \hbo 
v(x) = a^{-1} \, xy \; . 
\label{eq:6d} 
\en 

The direction of this drift force tangent to the gauge orbit 
is illustrated in figure \ref{fig:0} (right panel). 

\vskip 0.3cm 
Rather than seeking the solution of the Fokker Planck equation 
for the case (\ref{eq:6d}), I employ the corresponding 
Langevin equation to generate a ``time'' history $x_k^{(n)}$ of the 
vectors and to calculate expectation values. The ``time'' history 
is obtained by the recursion 
\bea
x^{(n+1)} &=& x^{(n)} \; + \; dt \; \biggl( - \frac{ \partial S_{YM}
}{ \partial x } \; + \; a^{-1} \, x \, y^2 \biggr) \; + \; \eta _x \; , 
\label{eq:7} \\ 
y^{(n+1)} &=& y^{(n)} \; + \; dt \; \biggl( - \frac{ \partial S_{YM}
}{ \partial y }  \; - \; a^{-1} \, x^2 \, y \biggr) \; + \; \eta _y \; , 
\label{eq:8} 
\ena 
where $\eta _{x/y}$ is a Gaussian noise with width $\sqrt{4 \, dt }$. 
An extrapolation $dt \rightarrow 0 $ must be performed when 
expectations values are calculated. 
For the present example, $N=10^7$ pairs $(x_n,y_n)$ are created 
for $dt_1=0.01$ and $dt_2=0.005$, respectively. The estimator,
$s(dt)$, of a desired observable is calculated and the corresponding 
statistical error $\delta s (dt)$ is estimated. The final error
$\epsilon $ comprises statistical as well as systematic errors due 
to the extrapolation $dt \rightarrow 0$. I used the following 
procedure: 
$$ 
s \; = \; \frac{1}{2} \Bigl( s(dt_1) \, + \, s(dt_2) \Bigr) \; , \hbo 
\delta s^2 \; = \; \delta s^2 (dt_1) \, + \, \delta s^2 (dt_2) \, + \, 
(s(dt_1)-s(dt_2))^2 \; . 
$$ 
For a smaller value of $dt$, one must choose a larger value of 
$N$ of configurations in order to explore the complete 
``configuration space''. Generically, the Langevin approach is 
by far inferior as e.g.~the standard heat bath approach.

\vskip 0.3cm 
Let us firstly study the ``gauge invariant'' operator $\langle r^2 
\rangle (a^{-1})$ as function of the Landau parameter $a$. 
The numerical result is shown in figure \ref{fig:1}. The error bars 
comprise statistical as well as the systematic error from the 
extrapolation $dt \rightarrow 0 $. One indeed finds that 
$\langle r^2\rangle $ is independent of  $a$ within 
numerical accuracy. I also checked that $\langle r^4\rangle $ 
is independent of of  $a$. 

\begin{figure}[t]
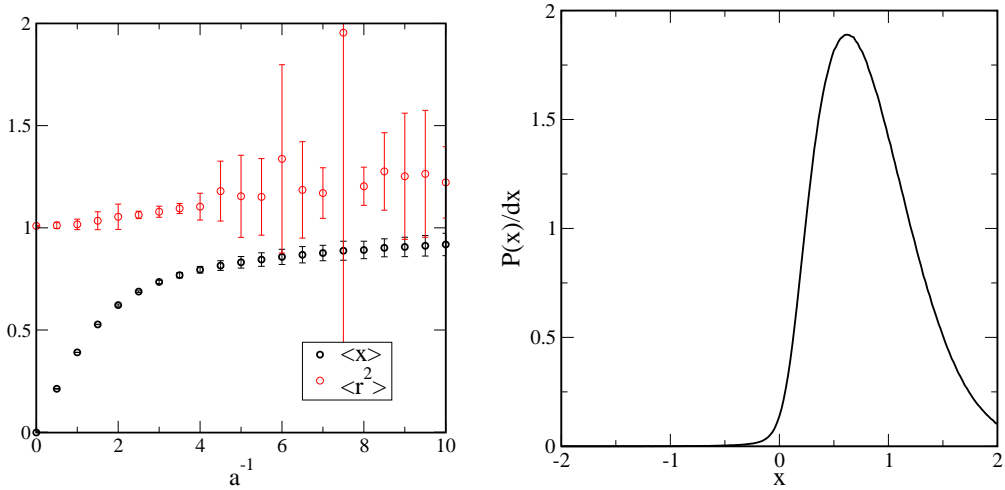

\centerline{  
\epsfxsize=6cm 
\epsffile{mod_all.eps} 
\hspace{0.3cm}
\epsfxsize=6.7cm 
\epsffile{modx.eps} 
} 
\vskip 0.3cm
\caption{ The expectation values $\langle r^2 \rangle $ 
   and $\langle y^2 \rangle $ 
   as function of the Landau parameter $a^{-1}$ (left panel). 
   The probability distribution of $x_n$ for $a^{-1}=5$ (right panel).  
   Drift force in (\ref{eq:10}).   
}
\label{fig:2}
\end{figure}
\vskip 0.3cm 
Secondly, I study the ``gauge variant'' expectation value 
$\langle y^2 \rangle $. The result is also shown in the above figure. 
One finds that $\langle y^2 \rangle = \langle r^2 \rangle /2 $  
for $a^{-1}=0$ as it should be. Increasing $a^{-1}$ decreases
fluctuations around $y=0$: the configurations $(x,y)$ are pushed 
to the first Gribov regime specified by $y=0$. I then 
calculated the expectation value of $\langle x \rangle$. It turns out 
that it vanishes for the investigated range of $a \in [0,10]$. 
This shows that there is still an average over the complete 
first Gribov regime, which consists of the positive {\it and } negative 
half $x$-axis. The probability distribution of $x$ is also shown 
in figure \ref{fig:1} for $a^{-1}=5$. 
The values $\vec{x}^{(n)}$ are symmetrically distributed 
around $x_1=x=0$. Hence, the configurations are sampled over the 
complete first Gribov regime (rather than the fundamental modular 
region, i.e., the positive half $x$-axis).

\vskip 0.3cm
{\bf Case study III. The fundamental modular region. } 

In this section, I will study a drift force which pushes the 
configurations $(x_n,y_n)$ towards the fundamental modular region 
$y=0$, $x \ge 0$. The drift force is given by 
\be 
K_i \; = \; \epsilon _{ik} x_k \, v(x) \; , \hbo 
v(x) = - a^{-1} \, y/r^3 \; , \hbo r = \sqrt{x^2+y^2}. 
\label{eq:10} 
\en 
The numerical simulation reveals that the ``gauge invariant'' 
observable $\langle r^2 \rangle $ is independent of the gauge 
fixing parameter $a^{-1}$. In addition, the points $(x^{(n)},y^{(n)})$ 
are nicely attracted by the positive half $x$-axis 
(see figure \ref{fig:2}). 

\vskip 0.3cm 
The crucial observation in deriving the function $P_v(x)$ is that 
the drift force (\ref{eq:10}) can be written as the gradient of a 
gauge fixing action function, i.e., 
\be 
K_i \; = \; - \; \partial _i S_\mathrm{fix} \; , \hbo 
S_\mathrm{fix} \; = \;  a^{-1} \; x/r \; . 
\label{eq:11} 
\en 
Hence, the Fokker-Planck equation is given by 
\be 
H_{FP} P_v \; = \; \frac{ \partial }{ \partial x_i } 
\biggl[ - \frac{ \partial P_v }{ \partial x_i }  \; + \; 
\Bigl( - \frac{ \partial S_{YM} }{ \partial x_i } \, - \,  
\frac{ \partial S_\mathrm{fix} }{ \partial x_i } \Bigr) P _v 
\; \biggr] \; = \; 0 \; . 
\label{eq:12} 
\en 
\begin{figure}[t]
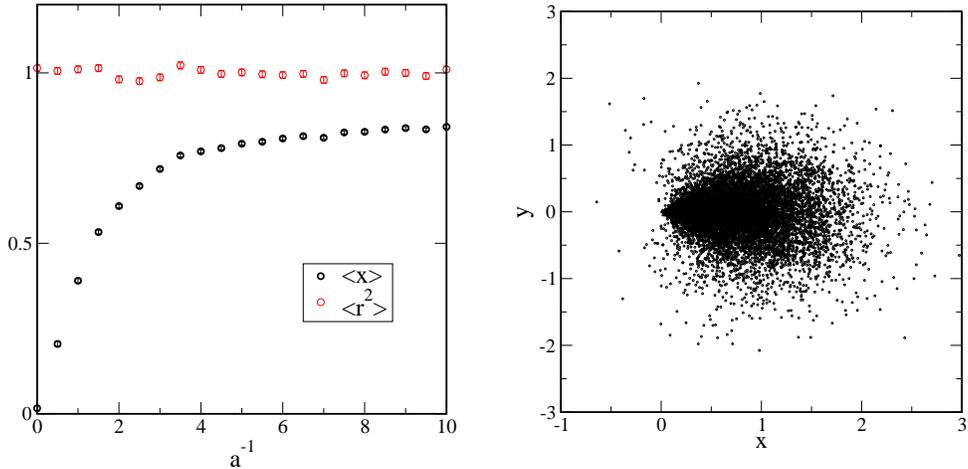

\centerline{  
\epsfxsize=6cm 
\epsffile{heat_all.eps} 
\hspace{0.5cm}
\epsfxsize=6cm 
\epsffile{scatter.eps} 
} 
\vskip 0.3cm
\caption{ The expectation values $\langle r^2 \rangle $ 
   and $\langle y^2 \rangle $ 
   as function of the Landau parameter $a^{-1}$  from a heat-bath 
   simulation (left panel). 
   The scatter plot of $10^4$ heat-bath configurations 
   for $a^{-1}=5$ (right panel); Drift force in (\ref{eq:10}).   
}
\label{fig:3}
\end{figure}
From this representation, one can read of the desired solution, i.e., 
\be 
P_v(x) \; = \; \exp \Bigl\{ - r^2 \; + \; a^{-1} \, \cos \varphi \, 
\Bigr\} \; . 
\label{eq:13} 
\en 
Here, the angular part and the radial part factorizes. This proves
that gauge invariant observables are indeed independent of the 
gauge fixing parameter. The maximum probability of $P_v(x)$ 
is obtained for $\varphi = 0$, which is the positive half  $x$-axis,
i.e., the fundamental modular region.

\vskip 0.3cm
{\bf The heat-bath simulation.} 

Finding the gauge fixing action $S_\mathrm{fix}$ which generates 
the drift force paves the path to an efficient simulation 
using the heat-bath approach. Thereby, configurations $(x,y)$ 
are generated according the probability 
\be 
\exp \Bigl\{ - S_{YM} \; - \; S_\mathrm{fix} \Bigr\} \; , 
\label{eq:14} 
\en 
where $S_\mathrm{fix} $ is given by (\ref{eq:11}). 
It turns out that $N_h=10^4$ heat-bath configurations are sufficient 
for error bars at the 1\% level (compare figure \ref{fig:2} with 
figure \ref{fig:3}). A scatter plot of the configurations $(x_n,y_n)$ 
is also shown in figure (\ref{eq:3}). 

\vskip 0.3cm
In conclusion, the heat-bath approach is roughly 
two orders of magnitude more efficient than the corresponding 
Langevin-approach. It is therefore highly desirable to construct 
the gauge fixing function $S_\mathrm{fix}$ generating the 
drift force. 

\vskip 0.3cm
For these purposes, 
the drift force $K_i$ must obey certain constraints in order to be 
generated by a gauge fixing action. These constraints are derived 
from (\ref{eq:11}) by observing that 
\bea 
- \; \Bigl( \partial _k \partial _i \; - \; \partial _i \partial _k 
\Bigr) \; S_\mathrm{fix} &=& 0 \; , 
\label{eq:15} \\ 
\partial _k K_i \; - \; \partial _i K_k &=& 0 \; .
\label{eq:16} 
\ena
The only non-trivial information is obtained from (\ref{eq:16}) 
by choosing $i=1$ and $k=2$. Using $K_i$ from (\ref{eq:10}), we find 
\be 
x \; \partial _x \; v(x,y) \; + \; y \; \partial _y \; v(x,y) 
\; + \; 2 \, v(x,y) \; = \; 0 \; . 
\label{eq:17} 
\en 
The last expression is most instructive using polar coordinates, and 
(\ref{eq:17}) becomes 
\be 
r \; \partial _r \; \ln \, \bar{v}(r,\varphi) \; + \; 2 \; = \; 0 \; .
\label{eq:18} 
\en 
Hence, the general solution of $\bar{v}(r,\varphi)$ which admit 
a representation in terms of a gauge fixing function 
$S_\mathrm{fix}$ is given by 
\be 
v(x,y) \; = \; 
\bar{v}(r,\varphi) \; = \; \frac{1}{r^2} \; f(\varphi ) \; , 
\label{eq:19} 
\en 
where the function $f(\varphi ) $ is still arbitrary.

\section{ Conclusions} 

A simple toy model was designed to illustrate the Gribov problem 
of the standard Faddeev Popov quantization. Using this model, 
the resolution of the Gribov problem along the lines of Zwanziger's 
version of the stochastic approach~\cite{Zwanziger:1981kg} has been 
outlined. In the Zwanziger approach, the gauged configurations are 
attracted by the first Gribov regime. Here, it turned out that the 
approach can be modified in order to obtain attraction by the 
fundamental modular region. Moreover, a integrable drift force 
towards the fundamental region was developed. It was therefore possible 
to construct the gauge fixing action which generates the drift 
force in the Langevin simulation. The latter construction was essential 
to perform a heat bath simulation. In the case of the toy model, 
it was observed that the heat bath method is more efficient than 
the Langevin technique.

\bigskip 
{\bf Acknowledgments: } 
I would like to thank Dan Zwanziger for encouragement and many
helpful discussions on gauge fixing and the stochastic approach. 
I am indebted to Oliver Schr\"oder for valuable comments
concerning, in particular, the construction of the conservative drift force. 
I thank R.~Alkofer and M.~Quandt for comments on the manuscript. 

\end{document}